\documentclass[12pt]{iopart}

\usepackage[utf8]{inputenc} 
\usepackage{float} 
\usepackage{enumitem}
\usepackage{bbold}
\usepackage[normalem]{ulem}
\usepackage{hyperref} 
\hypersetup{colorlinks=true, urlcolor=blue, citecolor=blue,linkcolor=blue}
\usepackage[all]{hypcap} 

\usepackage{graphicx}
\usepackage{dcolumn}
\usepackage{bm}
\usepackage{psfrag}
\usepackage{epsfig}
\usepackage{multirow}
\usepackage{color}
\usepackage{bbm}
\usepackage[FIGTOPCAP,raggedright,nooneline]{subfigure}
\usepackage{verbatim}
\usepackage{upgreek} 
\usepackage{tikz} 
\usepackage[mathscr]{euscript}

\newcommand{\T}[1]{\text{#1}}
\newcommand{\bra}[1]{\langle #1 |}
\newcommand{\ket}[1]{| #1 \rangle}

\newcommand{\ignore}[1]{}

\newcommand{\eqs}{Eqs.\,}
\newcommand{\fig}{Fig.\,}

\newcommand{\cf} {cf.~}

\newcommand{\eg} {e.g.~}
\newcommand{\rref} {Ref.\,}
\newcommand{\rrefs} {Refs.\,}

\newcommand{\PT}{$\mathcal{PT}$}

\usepackage{iopams}  

\expandafter\let\csname equation*\endcsname\relax

\expandafter\let\csname endequation*\endcsname\relax

\usepackage{amsmath}

\begin{document}
\bibliographystyle{unsrt}

\title{Quantum correlations in \PT-symmetric systems}

\author{Federico Roccati$^{1,*}$, Salvatore Lorenzo$^{1}$, G.~Massimo Palma$^{1,2}$, Gabriel T.~Landi$^{3}$, Matteo Brunelli$^{4}$, Francesco Ciccarello$^{1,2}$}

\address{$^{1}$Universit{\`a}  degli Studi di Palermo, Dipartimento di Fisica e Chimica -- Emilio Segr{\`e}, via Archirafi 36, I-90123 Palermo, Italy}
\address{$^{2}$NEST, Istituto Nanoscienze-CNR, Piazza S. Silvestro 12, 56127 Pisa, Italy}
\address{$^{3}$Instituto de Fısica, Universidade de Sao Paulo, CEP 05314-970, Sao Paulo, Sao Paulo, Brazil}
\address{$^{4}$Cavendish Laboratory, University of Cambridge, Cambridge CB3 0HE, United Kingdom}

\address{$^{*}$Corresponding author: federico.roccati@unipa.it}


\vspace{10pt}
\begin{indented}
\item[]September 2020
\end{indented}

\begin{abstract}
 We study the dynamics of correlations in a paradigmatic setup to observe \PT-symmetric physics: a pair of coupled oscillators, one subject to a gain one to a loss. Starting from a coherent state, quantum correlations (QCs) are created, despite the system being driven only incoherently, and can survive indefinitely. \PT~symmetry breaking is accompanied by non-zero stationary QCs. We link \PT~symmetry breaking to the long-time behavior of both total and QCs, which display different scalings in the \PT-broken/unbroken phase and at the exceptional point (EP). This is analytically shown and quantitatively explained in terms of entropy balance. The EP in particular stands out as the most classical configuration.
\end{abstract}

%
%
%
%
%

\section{Introduction}
The finding of non-Hermitian Hamiltonians with real eigenvalues~\cite{benderPRL1998} fueled widespread attention at a fundamental level, as well as in terms of potential applications~\cite{el-ganainyNP2018,fengNP2017,longhiEPL2017}. A major motivation comes from the experimental implementability of such Hamiltonians, especially in optics~\cite{ruterNP2010,regensburgerN2012,pengNP2014}. A prototypical example  (see~\fig1) is a pair of coupled oscillators, separately subject to either gain or loss. At the mean-field level, the modes evolve according to a Schr\"odinger-like equation featuring a non-Hermitian Hamiltonian $\cal H$ that enjoys parity-time (\PT) symmetry~\cite{benderPLA2010}. 

To date the vast majority of studies of such dynamics adopted a classical description (based on Maxwell's equations in all-optical setups), thus neglecting quantum noise. Recent works yet showed that a full quantum treatment (beyond mean field) can have major consequences~\cite{kepesidisNJP2016,lauNC2018,zhangAQ2018,wangPRA2019}, although the exploration of this quantum regime
is still in an early stage~\cite{schomerusPRL2010,yooPRA2011,agarwalPRA2012,vashahri-ghamsariPRA2017,longhiOL2018,vashahri-ghamsariPRA2019,Miranowicz-PRA2019,klauck2019observation,Zeno_PRA2020,jaramillo2020pt,stone_PRL2020}. {With regard to the potential exploitation of \PT-symmetric systems for quantum technologies},  a major obstacle is that gain and loss unavoidably introduce quantum noise, which is detrimental for 
quantum coherent phenomena---entanglement above all~\cite{nielsen2010}. In particular, the incoherent pumping due to the gain is unusual in quantum optics settings~\cite{braunsteinRMP2005}. This issue even motivated recent proposals to employ parametric driving in place of gain/loss to effectively model non-Hermitian systems~\cite{McDonaldPRX2018,wangPRA2019}.

Yet, in the last two decades, ``cheaper'' quantum resources have been discovered that put milder constraints on the necessary amount of quantum coherence. Among these is quantum {\it discord}, a form of quantum correlations (QCs) that can occur even in absence of entanglement~\cite{ollivierPRL2001,hendersonJPAMG2001}. This extended paradigm of QCs
has received huge attention for its potential of providing a quantum advantage in noisy environments~\cite{modiRMP2012}.
Remarkably, a very recent work reported the first experimental detection of such a form of QCs~\cite{caoPRL2020} in an {anti}-\PT-symmetric system featuring similarities with the setup in~\fig1. 
{However, the existence of a general connection between \PT~symmetry and QCs dynamics is yet unknown.}			
\begin{figure}
	\centering
	\includegraphics[width=0.7\columnwidth]{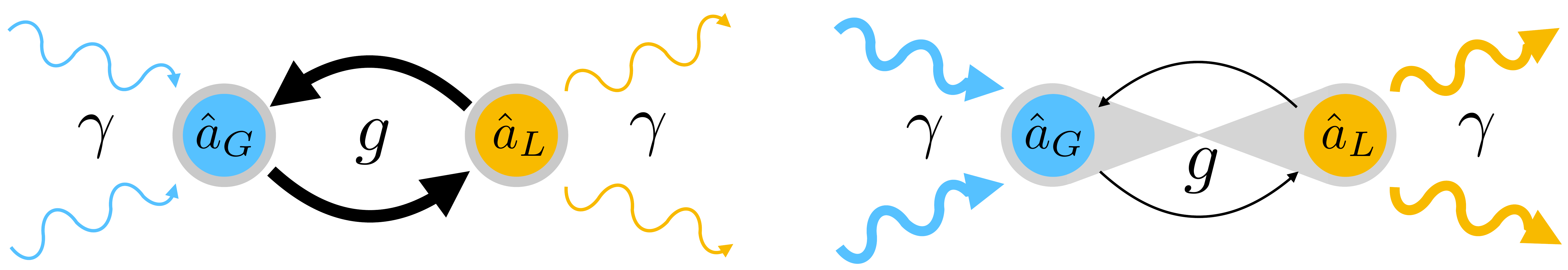}
	\caption{{A pair of quantum  oscillators $G$ and $L$ undergo a coherent exchange energy with rate $g$. Additionally, mode $G$ ($L$) is subject to a local gain (loss) with rate $\gamma$. The mean-field dynamics is described by a \PT-symmetric Hamiltonian. (Left): when \PT~symmetry is preserved ($g>\gamma$), if each mode starts in a coherent state (zero correlations), after some time they will share only classical correlations. (Right): \PT~symmetry breaking ($g<\gamma$) is instead accompanied by stationary quantum correlations.}} 
\end{figure}

This work puts forth a detailed study of total and quantum correlations in the case study of the gain-loss setup in~\fig1, which is the simplest and most widely investigated system to implement \PT-symmetric non-Hermitian Hamiltonians \cite{el-ganainyNP2018}. A link emerges between \PT~symmetry breaking and the long-time behavior of both total and QCs: these are found to display different scalings in the \PT-broken/unbroken phase and at the exceptional point (EP). This is proven analytically and the underlying mechanism explained in detail through entropic arguments. In particular, breaking of \PT~symmetry is accompanied by the appearance of finite stationary discord. 
Our study provides a new characterization of phases with unbroken/broken \PT~symmetry in terms of the asymptotic behavior of correlations, whose knowledge requires accounting for the full quantum nature of the field.

\section{System}
We consider two quantum harmonic oscillators $G$ and $L$ (see~\fig1), whose joint state 
evolves in time according to the Lindblad master equation (we set $\hbar=1$ throughout)
\begin{equation}
\dot\rho  =  - i[g( \hat a_{L }^\dagger \hat a_{G}+{\rm H.c.}) ,\rho] +2\gamma_{L}\, \mathscr D[\hat a_L]\rho +2\gamma_{G} \,\mathscr D[\hat a_G^\dag]\rho \label{ME}
\end{equation}
with 
\begin{equation}
\mathscr D[\hat A]\rho=\hat A \rho \hat A^\dagger - \frac{1}{2}(\hat A^\dagger \hat A \rho+\rho\hat A^\dagger\hat A).\nonumber 
\end{equation}
Here, $\hat a_{n}$ and $\hat a^\dag_{n}$ with $n=L,G$ are usual bosonic ladder operators $[\hat a_{n},\hat a^\dag_{n}]=1$ (we assumed  
a rotating frame so as to eliminate the free Hamiltonian term $\omega_0 (\hat a_G^\dag\hat a_G+\hat a_L^\dag\hat a_L)$, which does not affect any $G$-$L$ correlations).
The coupling Hamiltonian in~(\ref{ME}) describes a coherent energy exchange at rate $g$ between the modes. In addition, each oscillator interacts {incoherently} with a local environment: the one on $G$ pumps energy with characteristic rate $\gamma_G$ (gain) while that on $L$ absorbs energy with rate $\gamma_L$ (loss). This system can be implemented in a variety of ways \cite{el-ganainyNP2018}, including coupled waveguides \cite{ruterNP2010}, microcavities \cite{pengNP2014}, inductively-coupled LRC circuits \cite{RLC} and coupled pendula \cite{pendula}.


\begin{equation}\label{evolMeanVal}
\mathcal H=\left(
\begin{matrix} 
-i\gamma_{L } & g \\
g & i\gamma_{G} 
\end{matrix} 
\right)
\end{equation}
The non-Hermitian matrix $\cal H$ generally has two complex eigenvalues with associated non-orthogonal eigenstates. For $\gamma_L=\gamma_G=\gamma$ (the so-called {``\PT~line"}), $\cal H$ is invariant under \PT~symmetry, corresponding to a swap $G\leftrightarrow L$ combined with time reversal (complex conjugation). In this case, its eigenvalues are $\varepsilon_\pm=\pm \sqrt{g^2-\gamma^2}$.
These are real in the unbroken phase (UP) $\gamma<g$ and complex in the broken phase (BP) $\gamma>g$, coalescing at the exceptional point (EP) $\gamma=g$ where the corresponding eigenstates become parallel~\cite{el-ganainyNP2018}.
Equations analogous to~(\ref{ME}) for the full-quantum description of \PT-symmetric systems also appeared elsewhere (see \eg \rrefs\cite{dastPRA2014,longhi2019quantum}).

\section{Second-moment dynamics}

The two oscillators have an associated quantum uncertainty described by a $4\times 4$ covariance matrix. Introducing  quadratures $\hat x_n=\tfrac{1}{\sqrt{2}}(\hat a_n+\hat a_n^\dag)$ and $\hat p_n=\tfrac{i}{\sqrt{2}} (\hat a_n^\dagger - \hat a_n)$ (with $n = G,L$), we define the covariance matrix as $\sigma_{ij} = \langle \hat X_i \hat   X_j+\hat X_j \hat X_i\rangle -2 \langle \hat X_i \rangle \langle \hat X_j \rangle$, where $\hat X_i=( \hat x_L, \hat p_L,\hat x_G, \hat p_G)$~\cite{gardiner2004}.
Following a standard recipe~\cite{gardiner2004}, the master equation~\eqref{ME} implies a Lyapunov equation for the  covariance matrix: 
\begin{equation}\label{eqforsigma_Y}
\dot{\sigma} = Y\sigma +\sigma\, Y^{T} + 4 D 
\end{equation}
with
\begin{equation}\label{Ymatrix}
Y=\left(
\begin{matrix}
-\gamma_{L }  & 0 & 0 & g \\
0 & -\gamma_{L }  & -g & 0 \\
0 & g & \gamma_{G}  & 0 \\
-g & 0 & 0 & \gamma_{G} \\
\end{matrix} 
\right)
\end{equation}
and
$D=\tfrac{1}{2}\,{\rm diag}(\gamma_{L},\gamma_{L},\gamma_{G},\gamma_{G})\label{D}$. 
The dynamics generated by Eq.~\eqref{ME} is Gaussian preserving, hence a Gaussian initial state will remain so at any time. Thereby, the entire state is fully specified by the mean-field vector $\psi$ and the covariance matrix $\sigma$~\cite{ferraro2005, olivares2012quantum}.

\section{Correlation measures}
A measure of the \emph{total} amount of correlations between $\hat a_G$ and $\hat a_L$ is given by the mutual information $\mathcal I= S_{G}+S_{L}-S$, which is the difference between the sum of local entropies  $S_{L(G)}=-\T{Tr}(\rho_{L(G)}\log\rho_{L(G)})$, with $\rho_{L(G)}={\rm Tr}_{G(L)}\rho$, and the entropy of the joint system
$S=-\T{Tr}(\rho\log\rho)$~\cite{cover2006,nielsen2010}.
It immediately follows that ${\cal I}=0$ if and only if $\rho=\rho_L\otimes \rho_G$. 
Instead, the amount of QCs is measured by the so-called {\it quantum discord}~\cite{ollivierPRL2001,hendersonJPAMG2001,modiRMP2012} 		
\begin{equation}
\mathcal D_{LG}=S_{G}-S+\underset{\hat G_k}{\rm min}\,\,\sum_k p_k S(\rho_{L|k})\,,\label{d-def}
\end{equation}
where the minimization is over all possible quantum measurements $\{\hat G_k\}$ made on $G$. 
A measurement outcome indexed by $k$ collapses the joint system onto $\rho_{L|k}=\Tr_G(\hat G_k \rho)/p_k$ with probability $p_k$. 
Note that discord is in general asymmetric, i.e., $\mathcal D_{LG}\neq \mathcal D_{GL}$, which is the typical case for our system [see~\fig2(c)]. The difference $\mathcal{I} - \mathcal{D}_{LG}$ quantifies the maximum amount of information that can be extracted about $L$ only from local measurements on $G$. {Based on this,} discord captures QCs beyond entanglement, as it is in general nonzero for separable states~\cite{modiRMP2012}. {Hence, correlations between the modes are wholly classical only when both $\mathcal D_{LG}$ and $\mathcal D_{GL}$ vanish.}

For Gaussian states, the optimization in~\eqref{d-def} can be restricted to Gaussian measurements (Gaussian discord)~\cite{pirandolaPRL2014}, leading to a closed-form, albeit cumbersome, expression for ${\mathcal D}$~\cite{giordaPRL2010,adessoPRL2010}.  
In order to provide a simpler analytic expression we replace the von Neumann entropy by the {\it R{\'e}nyi-2} entropy $S(\varrho)=-\log \T{Tr}(\varrho^2)$ in each expression~\cite{adessoOSID2014}. 
For Gaussian states, it  has been shown that the choice of R{\'e}nyi-2 entropy leads to well-behaved correlation measures~\cite{renyi2PRL}. 
We however numerically checked that all of the results presented (in particular asymptotic scalings) are qualitatively unaffected if von Neumann entropy is used instead.
The fact that discord detects QCs more general than entanglement is condensed in a simple property: states such that $\mathcal D>\log 2$ are entangled ($\log 2\rightarrow$1 if von Neumann entropy is used)~\cite{adessoPRL2010}.

\begin{figure*}[t]
	\includegraphics[width=\linewidth]{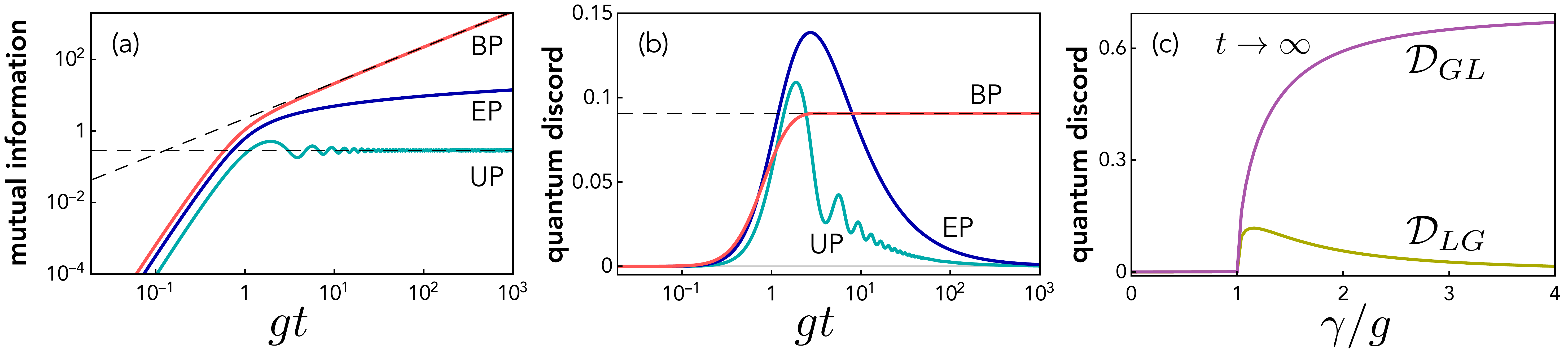}
	\caption{Evolution of total and quantum correlations on the \PT~line ($\gamma_{L }=\gamma_{G}=\gamma$). This comprises the unbroken phase (UP) $\gamma<g$, the exceptional point (EP) $\gamma=g$ and the broken phase (BP) for $\gamma>g$.
		(a) and (b): Mutual information $\cal I$ (a) and discord ${\cal D}_{LG}$ (b) for $\gamma=g/2$ (UP, green), $\gamma=3g/2$ (BP, red) and $\gamma = g$ (EP, blue). A qualitatively analogous behavior is exhibited by ${\cal D}_{GL}$. (c):  Asymptotic value of discord, $\mathcal{D}_{LG}(\infty)$ (yellow) and $\mathcal{D}_{GL}(\infty)$ (purple),
		~\cite{SM}
		} \label{fig2}
\end{figure*}

\section{Correlations dynamics for balanced gain and loss}
We study the dynamics of correlations when each oscillator $n=L,G$ starts in a coherent state $\ket{\alpha_{n}}=e^{(\alpha\hat a_n^\dag-\alpha^*\hat a_n)}\ket{0}$; the initial covariance matrix is thus simply $\sigma_0=\mathbb{1}_4$.
We evolve the covariance matrix through~\eqref{eqforsigma_Y} and then compute the time evolution of both correlation measures $\mathcal I$ and $\mathcal D$ (we verified that entanglement is indeed always zero), for which we obtain exact and compact expressions,~\cite{SM}.
In particular, \eqref{d-def} admits a global minimum for all possible parameter values, which corresponds to a phase-insensitive (heterodyne) measurement. Intuitively, this property can be traced back to the absence of any coherent drive: the dynamics in~\eqref{ME} preserves $U(1)$ symmetry and thus favors the conditioning of phase-insensitive measurements over phase-sensitive ones; this in turn makes the latter suboptimal for generating QCs.

\fig2  shows the typical time behavior of mutual information ${\cal I}$ (a) and discord ${\cal D}$ (b) in the UP (green line), at the EP (blue) and in the BP (red). 
{Correlations, including QCs, are created on a typical time scale (transient time)} of the order of $\sim g^{-1}$ or less,~\cite{SM}.
{As discussed later on, \emph{transient} generation of QCs is common in noise-driven multipartite systems. In the long-time limit, instead, correlations show a peculiar behavior, which we next analyze for each phase.}

In the UP, ${\cal I}$ saturates to a finite value and exhibits secondary oscillations at frequency 2$\sqrt{g^2{-}\gamma^2}$, while discord {slowly decays until it vanishes}.
Their asymptotic expressions are given by~\cite{SM} 
\begin{equation}
\mathcal I\approx\log(\tfrac{g^2}{g^2-\gamma^2}),\quad \mathcal D_{LG},\mathcal D_{GL}\approx \tfrac{\gamma}{2g^2t} \label{scaling-UP}\,,
\end{equation}
(throughout the symbol $\approx$ indicates the long-time limit)
showing that discord {undergoes a power-law decay} in this phase. 
Thus in the UP asymptotic correlations are \emph{entirely classical}, i.e., they do not involve any quantum superposition. {At a glance, this may seem to contradict the well-known property that Gaussian states such that ${\cal I}=0$ are all and only those with zero discord~\cite{adessoPRL2010}. That property yet holds for systems with bounded mean energy, while the present dynamics is {\it unstable}  on the whole \PT~line (see \rref\cite{SM} for details
)}. 

When \PT~symmetry is broken, on the other hand, the behavior of long-time correlations changes dramatically. The mutual information now  grows linearly as $\mathcal I \approx 2\Omega\,t$, with $\Omega=\sqrt{\gamma^2-g^2}$, while QCs tend to a finite value given by
\begin{equation}
\mathcal D_{LG}\approx \log \!\left(\tfrac{\gamma  (\gamma +\Omega )+g^2}{2 \gamma ^2}\right),\,\mathcal D_{GL}\approx \log\! \left(\tfrac{\gamma  (3 \gamma +\Omega )-g^2}{2 \gamma ^2}\right)\label{Dstat}.
\end{equation}
{Thus in the BP \emph{stationary} QCs are established, notwithstanding the noisy action of gain/loss and despite the dynamics being unstable.}

{Jointly taken, \eqs(\ref{scaling-UP}) and (\ref{Dstat}) show that the nature of long-time correlations is different in the two phases. In each phase, stationary finite correlations occur, but these are {purely classical} in the UP (where ${\cal I}$ converges, while ${\cal D}\rightarrow0$) and {quantum} in the BP.}

{Finally, a special behavior occurs at the EP with the correlations scaling as }
\begin{equation}
\mathcal I\approx\log(\tfrac{4g^2}{3}t^2),\quad \mathcal D_{LG},\mathcal D_{GL}\approx \tfrac{1}{gt} \label{scaling-EP}.
\end{equation}
{Thus, while discord scales as in the UP phase (although with a different pre-factor, \cf~\eqref{scaling-UP}), the growth of mutual information is now logarithmic. 
	Notably, the EP is the only point on the \PT~line such that $\mathcal{I}\rightarrow\infty,\,\mathcal{D}\rightarrow0$ (purely classical and diverging correlations). Thus, for balanced gain and loss, the EP can be regarded as the \emph{most classical} configuration.}

\fig2(c) shows the stationary QCs on the \PT~line. In the BP, ${\mathcal{D}}_{GL}(\infty)$ monotonically grows with $\gamma$ {asymptotically approaching the entanglement threshold, while ${\mathcal{D}}_{LG}(\infty)$} takes a maximum followed by a long-tail decay. 
{A critical behavior occurs at the EP (on the boundary between regions of zero and non-zero discord) since ${\mathcal D}\sim (\gamma/g-1)^{\frac{1}{2}}$ for $\gamma>g$ while ${\mathcal D}=0$ for $\gamma\le g$.}

As specified previously, all the plots in \fig2 are for an initial coherent state. Yet, we gathered numerical evidence that different initial Gaussian states yield analogous long-time behaviors of correlations~\cite{SM}
for details.

\section{Physical mechanisms behind generation of correlations}
{Generation of QCs} during the {\it transient} dynamics can be understood by noting that
the coupling Hamiltonian acts on the modes like a beam splitter. When acting on $\ket{\alpha_L}\otimes \ket{\alpha_G}$,
this term alone  cannot correlate the modes, only mixing their amplitudes~\cite{kimPRA2002}. 
The same is also true of the loss term. 
The gain, on the other hand, turns a coherent state into a {\it mixture}  
\begin{equation*}
\ket{\alpha}\!\bra{\alpha}\rightarrow \int d^2 \alpha'P(\alpha') \ket{\alpha'}\!\bra{\alpha'}
\end{equation*} 
with $P(\alpha')\ge 0$ (coherence reduced)~\cite{scheelEPL2018}. The combined action of gain and beam splitter on $\ket{\alpha_L}\!\bra{\alpha_L}\otimes \ket{\alpha_G}\!\bra{\alpha_G}$ turns it into 
\begin{equation*}
\int d^2\alpha'_G P(\alpha'_G) \ket{\tilde\alpha'_L}\!\bra{\tilde \alpha'_L}\otimes\ket{\tilde\alpha'_G}\!\bra{\tilde \alpha'_G}
\end{equation*}
where both $\tilde\alpha'_{L(G)}$ depend on both $\alpha'_{L(G)}$. Although disentangled, one such state is generally discordant because coherent states form a {\it non}-orthogonal basis~\cite{korolkovaRPP2019}. We note that a similar effect is obtained if the gain is replaced by a local thermal bath. Indeed, the ability of certain local non-unitary channels to favor creation of discord was demonstrated in~\cite{modiRMP2012}. For instance, local gain or loss can create QCs starting from a state featuring only classical correlations (a  process which is not possible for entanglement)~\cite{ciccarelloPRA2012a,huPRA2012,streltsovPRL2011,ciccarelloPRA2012}, which was experimentally confirmed in Ref.~\cite{madsenPRL2012}.

\begin{table}
	\centering
	\includegraphics[width=0.65\textwidth]{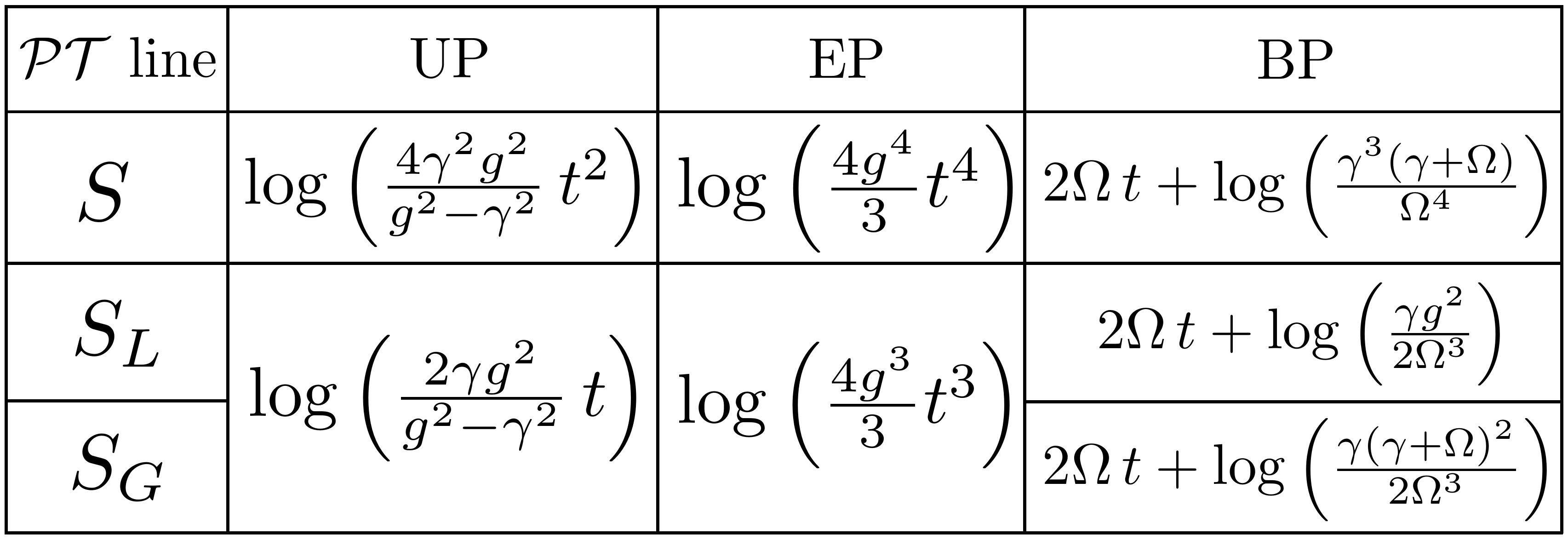}
	\caption{Asymptotic behavior of $S$ and $S_{L(G)}$ on the \PT~line.} 
	\label{table}
\end{table}

The peculiar nature of the present dynamics mostly comes from the {\it long-time} behavior of correlations. 
{To shed light on it, we first express the discord in the form }
\begin{equation}
\mathcal D_{LG} = \log\! \left(1+\frac{e^{\mathcal I}-1}{e^{S_G}+1}\right)\label{central}\,,
\end{equation}
(with an analogous expression  for $L\leftrightarrow G$). This identity{, which is proven in \rref~\cite{SM},
holds true} for \emph{any} Gaussian state generated by~\eqref{ME} and subject to a local heterodyne measurement.
{Combined with ${\cal I}=S_G+S_L-S$, \eqref{central} allows to explain the dynamics of classical and QCs in terms of a \emph{competition} between global and local entropies.}  {The long-time expressions of $S$ and $S_{L(G)}$ are reported in Table~\ref{table}. All of these {\it diverge} in time (either logarithmically or linearly depending on the phase).
	Hence, \eqref{central} simplifies to}
\begin{equation}
\mathcal D_{LG} \approx  \log \left(1 + e^{-(S-S_L) }-e^{-S_G}\right)\,,\label{approx}
\end{equation}
which shows that the survival of QCs is controlled by $S-S_L$ alone. {Using the expressions in Table~\ref{table}, \eqref{approx} yields precisely the scalings in \eqs(\ref{scaling-UP}), (\ref{Dstat}) and (\ref{scaling-EP}). }

In the UP,  
$S_G$ and $S_L$ grow at the same rate and their sum is \emph{almost} equal to the global entropy $S$. Their difference is small ({showing this requires  sub-leading contributions not reported in} Table~\ref{table}) and yields constant ${\cal I}$ in the long-time limit. {\eqref{central} then entails that discord vanishes}.
In the BP, instead, the gain dominates the entropy balance and the total entropy is slaved to the local one, $S\approx S_G$. This in turn implies $\mathcal{I}\approx S_L$. 
Moreover, the divergences of $S$ and $S_L$ cancel out, so that $S-S_L$ is convergent, in turn entailing a finite value of QCs via \eqref{approx}.

{As mentioned previously, any two-mode Gaussian state with finite mean energy fulfills ${\cal D}\neq 0\Leftrightarrow{\cal I}\neq 0$~\cite{adessoPRL2010}}. This property can be retrieved from \eqref{central} when $S_G$ is {\it finite}. Yet, for $S_G\rightarrow \infty$, discord can vanish asymptotically even if ${\cal I}$ does not (\eg in the UP and at EP, see~\fig2).

\begin{figure}
	\centering
	\includegraphics[width=0.7\columnwidth]{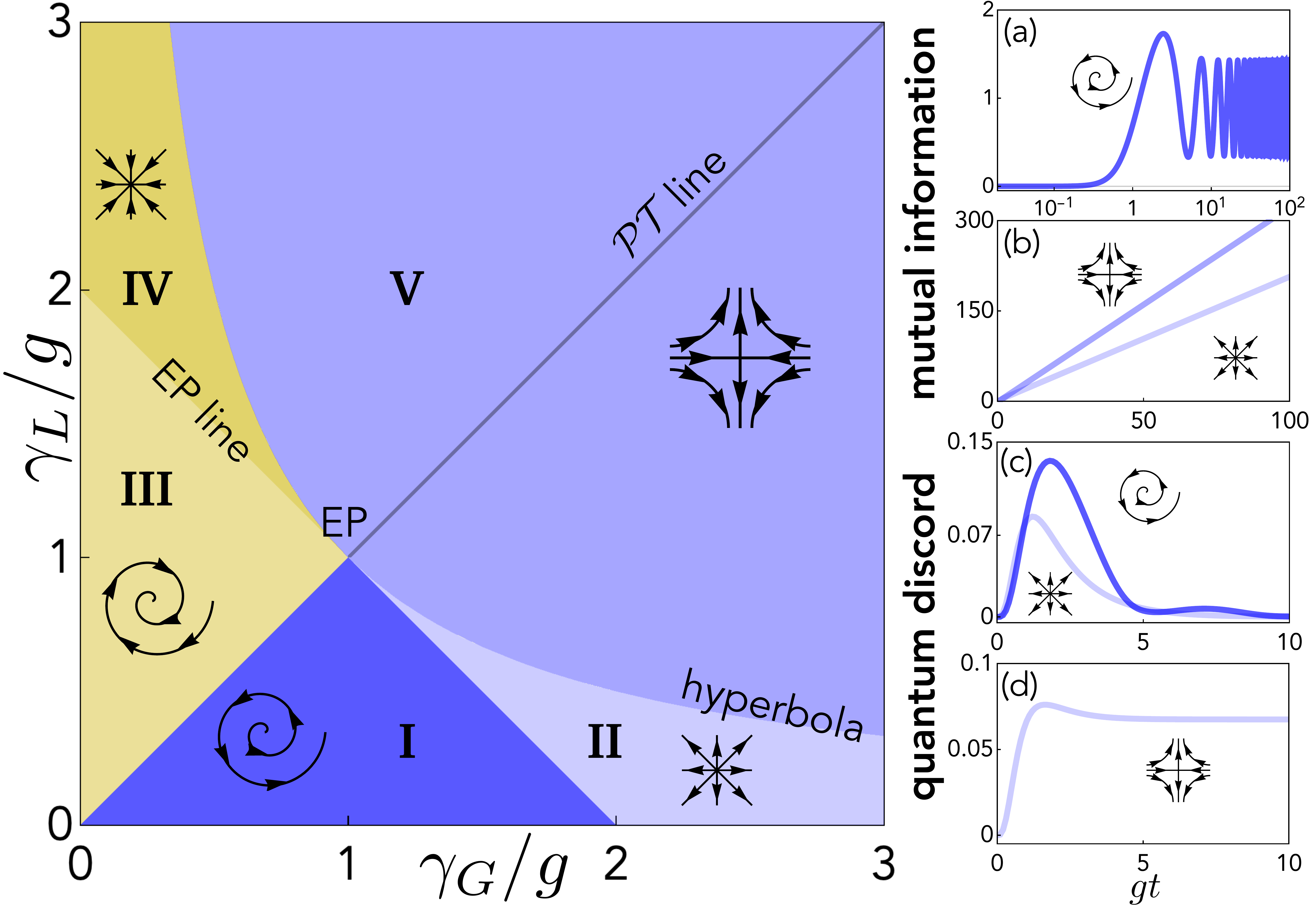}
	\caption{(Left): Stability diagram for the dynamics ruled by \eqref{ME} with unbalanced gain and loss; see text for details. (Right): {Typical time evolution of the mutual information} [(a), (b)] and discord [(c), (d)] corresponding to the points highlighted in the diagram.} 
\end{figure}

\section{Dynamics of correlations beyond the PT~line}
Lastly, we address the  rich dynamics of correlations beyond the \PT~line, i.e.,~for unbalanced gain and loss ($\gamma_L\neq \gamma_G$). The phase portrait in~\fig3 displays five distinct dynamical regimes, obtained by applying standard stability analysis {(see~\rref\cite{SM}
for details)}. These regions are limited by the \PT~line, the EP line $\gamma_L+\gamma_G=2 g$ and the hyperbola $\gamma_L\gamma_G=g^2$. There is a stable region (III+IV), {where both distinct eigenvalues $\lambda_{\pm}$ of matrix $Y$ (\cf\eqref{Ymatrix}) have negative real part (note that for $g>\gamma_G$ a too large rate $\gamma_L$ makes the dynamics unstable)}. This is the usual bounded-energy region featuring non-zero stationary values of $\mathcal{I}$ and $\mathcal{D}$. Symmetric to that is a totally unstable region (I+II), where both $\Re \lambda_\pm>0$. Notably, this \emph{whole} region is characterized by asymptotically vanishing discord [\cf\fig3(c)]. 
The EP line separates two kinds of divergence (convergence) in the totally unstable (stable) region: {below this line there occur repulsive (attractive) spirals, and sources (sinks) above it. }
Finally, there is an unstable region (V) (saddle points) with linearly divergent $\mathcal{I}$ and stationary QCs [\cf(b) and (c)]. 
\eqref{approx} can be directly applied to the unstable regions beside the \PT~line to explain the behavior of QCs.
Yet another remarkable feature is that the region (I+III+IV) is characterized by asymptotic finite values of $\mathcal{I}$. In particular, in region I, $\mathcal{I}$ displays extremely long-lived oscillations [see~\fig3(a)].

\section{Conclusions}
Through a fully quantum description, we studied the dynamics of total and quantum correlations in a typical gain-loss system exhibiting \PT-symmetric physics. With the modes initially in a coherent state, QCs without entanglement are created and, in a large region of the parameter space, settle to a non-zero value. 
For balanced gain and loss and in the long-time limit, phases with distinct \PT~symmetry exhibit dramatically different time scalings of both total and quantum correlations. This suggests a new distinction between phases with unbroken/broken \PT~symmetry in the dynamics of entropic quantities, whose knowledge requires accounting for the full quantum nature of the field. 
From the viewpoint of QCs theory, the unstable nature of the dynamics brings about exotic behaviors such as diverging correlations of a purely classical nature, which arise at the exceptional point. In terms of quantum technologies, stationary QCs beyond entanglement (occurring e.g.~in the unbroken phase) are potentially appealing in that this form of correlations have found several applications in recent years,~\cite{adessoJPAMT2016,streltsov2015} such as information encoding~\cite{guNP2012}, remote-state preparation~\cite{dakicNP2012}, entanglement activation~\cite{pianiPRL2011, adessoPRL2014, streltsovPRL2011a, croalPRL2015}, entanglement distribution~\cite{chuanPRL2012,peuntingerPRL2013, vollmerPRL2013, fedrizziPRL2013}, quantum metrology and sensing~\cite{girolamiPRL2014} and so on.
This suggests that quantum noise could embody a resource, rather than a hindrance, to the exploitation of \PT-symmetric systems for useful applications. Future important tasks will be studying the effect of finite temperature and gain saturation (the latter introduces non-linearities affecting the Gaussian nature of the dynamics).

\vspace{1cm}

\noindent \textbf{Acknowledgements.} We acknowledge fruitful discussions with 
M.~Paternostro, V.~Giovannetti, P.~Rabl, and T.~Tufarelli. F.~R.~acknowledges partial support from GNFM-INdAM.  G.~T.~L.~acknowledges the University of Palermo, for both the hospitality and the financial support. M.~B.~acknowledges support by the European Union Horizon 2020 research and innovation programme under grant agreement No 732894 (FET Proactive HOT).

\vspace{1cm}

\section*{References}

\bibliography{all}

\end{document}